\renewcommand{\d}{\delta}

\newcommand{\cD}{{\cal D}}

\newcommand{\ep}{\epsilon}

\newcommand{\si}{\sigma}
\newcommand{\Si}{\Sigma}

\newcommand{\p}{\partial}

\newcommand{\beq}{\begin{equation}}
\newcommand{\eeq}{\end{equation}}
\newcommand{\beqa}{\begin{eqnarray}}
\newcommand{\eeqa}{\end{eqnarray}}

\newcommand{\prd}[1]{{ \it Phys.~Rev.}~{\bf D{#1}}}
\newcommand{\prl}[1]{{ \it Phys.~Rev.~Lett.}~{\bf {#1}}}
\newcommand{\plb}[1]{{ \it Phys.~Lett.}~{\bf {#1B}}}
\newcommand{\pla}[1]{{ \it Phys.~Lett.}~{\bf {#1A}}}
\newcommand{\npb}[1]{{ \it Nucl.~Phys.}~{\bf B{#1}}}

\documentstyle[12pt,epsf]{article}
\newcommand{\postscript}[2]
 {\setlength{\epsfxsize}{#2\hsize}
  \centerline{\epsfbox{#1}}}
\def\llap#1{\hbox to 0pt{\hss#1}}
\def\pola{a\llap{\hbox{\char'30\kern-1.2pt}}}
\def\pole{e\llap{\hbox{\char'30\kern-.8pt}}}

\begin{document}
\baselineskip 22pt plus 2pt

\begin{titlepage}
\renewcommand{\thefootnote}{\fnsymbol{footnote}}
\begin{flushright}
\parbox{1.in}
{
 IFT-1/94\\
 January 1994}
\end{flushright}
\vspace*{.5in}
\begin{centering}
{\Large TOWARD QCD STRING: NO FOLDS. \footnote{
Work supported, in part,
by Polish Government Research Grant KBN 2 0165 91 01 and KBN 2 0417 91 01.
}}\\
\vspace{2cm}
{\large        Jacek Pawe\l czyk}\\
\vspace{.5cm}
        {\sl Institute of Theoretical Physics, Warsaw University,\\
        Ho\.{z}a 69, PL-00-681 Warsaw, Poland.}\\
\vspace{.5in}
\end{centering}
\begin{abstract}
We propose a string theory model which explains
several features of two dimensional YM theory.
Folds are suppressed.
This in turn leads to the empty theory in flat target spaces.
The Nambu-Goto action appears in the usual  way.
The model naturally  splits  into
two (chiral) sectors: orientation preserving maps and
orientation reversing maps.
Moreover it has a straightforward extension to 3 and 4 dimensional
space-times, which could be the rigid string with the
self-intersection number at $\theta=\pi$.

\end{abstract}

\end{titlepage}

\setcounter{footnote}{0}
\renewcommand{\thefootnote}{\arabic{footnote}}

It is strongly believed that the dynamics of gauge fields can be described
in terms of a string theory. The idea was supported by the lattice strong
coupling expansion \cite{lat} and the $1/N_c$  expansion \cite{thooft}. The
latter applied
in 2-dimensional (2D) models gave several well established relations
between QCD$_2$ (or YM$_2$) and a string theory \cite{bars,gross}. It
appeared that the crucial role !
is played by  the no fold condition, which strongly restricts the set of
the surface-to-surface maps which defines the string theory. Moreover, the
results of \cite{bars,gross}  indicate that the proper string action
should contain
the Nambu-Goto term.  It is well known that the Nambu-Goto
term alone can not give the correct
picture, because the appropriate functional integral can not suppress
folds.

In this note we propose a resolution to this
problem. We shall supplement the Nambu-Goto
action by a topological term which then will lead to cancellation of folds.
The topological term is well defined in a target space-time of dimension 4 so
we shall introduce two additional variables with values in $R^2$.
The new variables will enter only the topological term - in this sense they
will not have  any dynamics.
We shall show that the model
has null partition function and null transition amplitudes for
microscopic states (infinitesimal punctures) for string propagating on the flat
2D target space-time.

The topological term we are going to consider is the self-intersection
number $(I)$ \cite{whitney1} of a surface
immersed in the extended  4-dimensional space-time.
It has been previously considered in \cite{bal,mazur,moja}.
It is a topological invariant in
some respects, similar to $F{\tilde F}$ of  YM${}_4$: for example, the string
analog of the $U_A(1)$
anomaly
is proportional to $I$ \cite{mazur,moja}. In addition it may play a role in
giving  spin degrees of freedom to string ends where
we expect that quarks live \cite{moja}.

We depart from the Polyakov picture of the string theory \cite{pols}
in the sense that no dependence on the elementary world-sheet
metric is involved -
the only world-sheet metric we use is the
induced one. It is worth to note that the area-preserving
diffeomorphism plays no  role in the whole construction.

The string theory functional integral for 2D targets is a sum over
surface-to-surface maps $\Sigma\to M$, where $\Sigma$ denotes the string
world-sheet and
$M$ the 2D target space-time.
It is known that generic surface-to-surface maps
contain singularities which are folds and cusps \cite{whitney2}.
Hereafter we are going to consider only such maps.
Folds form a submanifold
$\Si$ for which
one eigenvalue of the induced metric $g_{ab}=\p_a {\vec X}\p_b {\vec X}$
vanishes.
In order to remove these singularities  we
introduce two additional, (vertical) world-sheet fields: $(X^3,\;X^4)\in
R^2$ and
consider all lifts of  the ($X^1,\;X^2$) map i.e.  maps of the form
$(X^1,\;X^2,\;X^3,\;X^4)\in M\times R^2$, where fields $(X^3,\;X^4)$ take
arbitrary values.  A lift is an
immersion, in general. It means that the image of $\Sigma$ has two linearly
independent tangent vectors so e.g. the induced metric is non-degenerate.
Folds are places where the surface-to-surface map cease to be an immersion.

The proposed string action is
\beq
S[X]=\mu\int_M d^2\si\sqrt{g}+i\theta I[X],
\label{arean}
\eeq
where $I$ is the self-intersection number of the  surface immersed
in the 4D space.
For topologically trivial target spaces it equals:
\beq
I=-\frac{1}{16\pi}\int_M \,d^2\xi\sqrt{g}g^{ab}\p_a t_{\mu\nu}
\p_b {\tilde t}^{\mu\nu}
\eeq
where ${\vec X}={\vec X}(\xi)$ defines the immersion and
$t_{\mu\nu}=\ep^{ab}\;\p_a X^\mu \p_b X^\nu /\sqrt{g}$.
The vertical coordinates  enter the action only through $I$.

The action (\ref{arean}) is invariant
under arbitrary v-regular homotopy of vertical fields:
$\d X^\mu(\xi)=\ep^\mu(\xi),\quad (\mu=3,\; 4)$ \cite{whitney1}. A
v-regular homotopy is
a homotopy (i.e. just mentioned shift symmetry) which is an immersion
for each homotopy parameter.

In the following we  are going to classify  topological sectors of the
model.
We say that two immersions are in the same topological sector if they can
be connected by a v-regular homotopy.
Consider a 2D surface  with folds and its lift
into 4D space ($X^1, \dots, \;X^4$). The claim is that to
any fold (in ($X^1,X^2$) space) corresponds infinitely many
topologically inequivalent lifts characterized by a set of integer numbers.
The latter are assigned to folds and are invariant under
v-regular homotopies. A sum of these numbers
gives the self-intersection number $I\in Z$. We stress again that
all lifts have
the same dependence on $X^1,\;X^2$, e.g. the same area
(the same value  of the Nambu-Goto part of the action (\ref{arean})).

Consider a map of a sphere $S^2$ on $R^2$ with one
 fold $S_1$ (see Fig.\ 1 ). Thus the map is 2 to 1 everywhere except the fold
$S_1$
itself.
\begin{figure}[h]
\vspace{0.5cm}
\postscript{fig.eps}{0.87}
\vspace{0.5cm}
\caption{A map of  $S^2$ on $R^2$ with one
 fold $S_1$}
\end{figure}
The fold is topologically a circle $S^1$.
One can always choose one non-vanishing, continuous tangent
vector along the fold. It is the non-zero eigenvalue of the induced metric.
Lifts of the fold must have non degenerate 2D tangent space.
Hence, the another vector tangent to the immersed surface
has to lie in the ($X^3,\;X^4$) plane. All the possible lifts
belong to homotopy classes
of the maps $S^1\to S^1$ i.e. $\pi_1(S^1)=Z$.
The first $S^1$ is the fold, the second $S^1$ represents non-zero
tangents in  the  ($X^3,\;X^4$) plane. If $\sigma$ parameterize the fold
then
for given lift ($X^3(\sigma),\;X^4(\sigma)$) the element of $\pi_1(S^1)$ is
given
by the rotation number of this tangent: $f={1}/{2\pi} \int
d\sigma\;(\partial_\sigma
X^3  \partial_\sigma^2 X^4 - \partial_\sigma X^4 \partial_\sigma^2 X^3)
/((\partial_\sigma X^3)^2+(\partial_\sigma X^4)^2)$.
The integer $f$ is invariant under the v-regular homotopy and is directly
related to the self-intersection number $I$ of
the lifted configuration.
We can see it if we notice that both numbers are additive under
gluing. By gluing we mean a procedure of cutting  small
discs in both immersions and then connecting them by a tube.
Let us associate a pair $(f,I)$ of numbers to a lift.
If we glue it with  the $(f',I')$  lift, we obtain the
$(f+f',I+I')$ lift. Thus gluing $f_1$ copies of the $(-1,-I_1)$  lift with
the $(f_1,1)$  lift we get the $(0,1-f_1 I_1)$  lift.
But $f=0$ corresponds to $I=0$, because the above simple map with $f=0$ can
be lifted to an immersion in just 3D space instead of 4D space.
Thus $1=f_1 I_1$, so $I_1=f_1=\pm 1$ (the sign is undetermined).
The same reasoning can be repeated
for more complicated folds with several disconnected components and
cusps. One assigns the numbers $f_i\in \pi_1(S^1)$ to $i$-th
connected component of the set of folds. The numbers $\{f_i\}$
are  invariant under the v-regular homotopy. The self-intersection number is
then $I[f]=\sum_{folds}\pm f_i$. One can see it gluing lifts of the just
considered  map with one fold to lifts of the other folds.

Below we give a simple proof of the following statement.

{\it V-regular homotopy classes of lifts (i.e. topological sectors of the
model) are in
one-to-one correspondence with sets $\{f_i\}$}.

Let us consider two lifts
$(X^1,X^2,X_1^3,X_1^4)$ and $(X^1,X^2,X_2^3,X_2^4)$.
If they are v-regularly homotopic then they define the same set $\{f_i\}$,
because $\{f_i\}$ is invariant under any v-regular homotopy.
On the other hand let us assume that
two immersions are characterized by the same set $\{ f_i \}$. Let $S_1$
denote folds of the map $(X^1,X^2)$.
The assumption implies that tangents to both lifts at $S_1$ are equal
up to a
v-regular homotopy - here it is a
local rotation of the tangents. Moreover, one can shift the lifts in such a
way
that they will be equal ($(X_1^3,X_1^4)=(X_2^3,X_2^4)$)  at $S_1$. Hence
both lifts are v-regularly homotopic
in an infinitesimally small  neighborhood of $S_1$ (up to second power
of an infinitesimal quantity). Away from the folds the
map  $(X^1,X^2)$ is an immersion. In this case the
shift parameters $\epsilon^\mu(\xi)$ ($\mu=3,4$) can be arbitrary. Thus
both lifts are v-regularly homotopic everywhere.

Now we go to the string theory. We want to show that the originally folded
 configurations $(X^1, X^2)$ will cancel out from the partition
function. Fixing the shift  symmetry in such a
way that the gauge slice picks only one
representant (or at least equal number of them)
of each topological sector we get the following expression
for the functional integral:
\beq
\int \cD X^1 \cD X^2\;e^{-S[X]}\sum_{\{f_i\}}e^{i \theta I[f]}
\eeq
The sum over $f_i$'s can be performed independently for each $i$ because
$I[f]=\sum_{folds}\pm f_i$. For one fold we get
\begin{equation}
\sum_{f\in Z}e^{\pm i\theta f}=2\pi\delta (\theta)
\label{sum}
\eeq
Thus all folded configurations vanish from the path integral for
non-zero $\theta$. Of course $\theta=\pi$ is the preferred value because
then the model does not break parity.

Maps contributing to the  vacuum-to-vacuum amplitude of the closed string
necessarily have folds for the target space $R^2$. According to the
above discussion the amplitude vanishes. This also holds for any correlation
function of any finite set of local operators.
Thus the final conclusion of this part of the paper is that
the model (\ref{arean}) is trivial for the $R^2$ space-time.

In the end let us discuss shortly the Nambu-Goto action for maps without
folds
which may occur for topologically non-trivial space-times.
In this case  one can distinguish two classes of maps:
orientation preserving ($+$) and
orientation reversing maps ($-$).
\beq
\int_M d^2\si \sqrt{g}=\pm \frac{1}{2}\int_M d^2\si
\ep^{\mu\nu}\ep^{ab}
\p_a X^\mu \p_b X^\nu, \quad \mu,\;\nu=1,2.
\label{vol}
\eeq
The sign is chosen in such a way that the r.h.s. is
positive. In this way the Nambu-Goto
action has been split into two (chiral) sectors of \cite{gross}.
To some extend both sectors can be considered separately.
In fact, this kind of
theory has been considered recently in \cite{moore} yielding
very interesting results.

If one views (\ref{arean}) as a compactification of
a 4D string then the additional dimensions should have topology of a
2D torus. We believe that the cancellation of folds holds also in this case
although  we have not proved it.
We must also stress that we can say very little  about non-generic
surface-to-surface maps.
Various subsets of these maps may be  relevant for the description of
various 2D gauge theories  \cite{witten,ya}. This will be considered in a
separate publication. It is clear that
the model (\ref{arean}) has straightforward extensions to 3 and 4
dimensional space-times.
It is enough to make the additional dimensions dynamical i.e.
add them to the Nambu-Goto action. Higher dimensional string models
may require more
terms e.g. the extrinsic curvature term \cite{polrig}.

\vskip1cm

{\bf Acknowledgment}. I would like to thank T. Mostowski for
discussions concerning geometrical aspects of the paper,
K.Gaw\pole dzki and
R.Dijkgraaf for reading the manuscript and valuable comments.

\end{document}